\documentclass[prb,twocolumn,showpacs]{revtex4}
\usepackage{graphicx} 
\usepackage{amsmath,amsthm,amssymb,mathrsfs}
\usepackage{bm}

\begin{document}

\title{Phenomenological spin transport theory driven by anomalous Nernst effect}

\author{Tomohiro Taniguchi}
 \affiliation{
 National Institute of Advanced Industrial Science and Technology (AIST), Spintronics Research Center, Tsukuba, Ibaraki 305-8568, Japan 
 }

 \begin{abstract}
{ 
Several experimental efforts such as material investigation and structure improvement 
have been made recently to find a large anomalous Nernst effect in ferromagnetic metals. 
Here, we develop a theory of spin transport driven by the anomalous Nernst effect in a diffusive ferromagnetic/nonmagnetic multilayer. 
Starting from a phenomenological formula of a spin-dependent electric current, 
the theoretical formulas of electric voltage and spin torque generated by the anomalous Nernst effect are derived. 
The magnitude of the electric voltage generated from the spin current via the inverse spin Hall effect is on the order of $0.1$ $\mu$V for currently available experimental parameter values. 
The temperature gradient necessary to switch the magnetization is quite larger than the typical experimental value. 
The separation of the contributions of the Seebeck and transverse spin Seebeck effects is also discussed. 
}
 \end{abstract}

 \maketitle




\section{Introduction}
\label{sec:Introduction}

The generation of spin current is a central topic in spintronics. 
An application of electric current to a current-perpendicular-to-plane (CPP) structure has been the most conventional method to generate spin current. 
The spin current interacts with the magnetizations in ferromagnets, and provides interesting phenomena 
such as magnetoresistance effect and spin torque excited magnetization dynamics \cite{baibich88,binasch89,pratt91,slonczewski96,berger96}. 
Alternative methods to generate spin current, 
such as nonlocal spin-injection \cite{johnson88,johnson93,jedema03,kimura06}, 
spin pumping by ferromagnetic resonance \cite{silsbee79,mizukami02a,mizukami02b,tserkovnyak02a,tserkovnyak02b}, 
spin Hall effect by spin-orbit interaction \cite{dyakonov71,hirsch99,kato04,ando08}, 
spin Seebeck effect by heating \cite{uchida08,uchida10,slachter10,bauer12}, 
and spin hydrodynamic generation by fluid dynamics \cite{takahashi16}, 
have also been proposed theoretically and demonstrated experimentally. 


A temperature gradient applied to a ferromagnet causes not only the Seebeck effect but also the anomalous Nernst effect, 
where an electric field is produced along the direction normal to the temperature gradient and the magnetization \cite{huang11,sakuraba13,sakuraba16,uchida15}. 
The magnitude of the anomalous Nernst effect ($\sim 1$ $\mu$V/K) 
is currently smaller than that of the Seebeck effect ($\sim 10$ $\mu$V/K).  
Recently, however, several efforts, such as  material investigation \cite{sakuraba13,sakuraba16} and structure improvement \cite{uchida15}, 
have been made to investigate a large anomalous Nernst effect in ferromagnetic metals. 
The motivation of these works is generating large electric power by simpler structure than the spin Seebeck system \cite{sakuraba13}. 
Note that the anomalous Nernst effect also generates spin current because of the spin-dependent transport properties. 
Note also that one can find a geometrical analogy between the anomalous Nernst effect and the anomalous Hall effect, 
which generates the electric field along the direction normal to the external electric voltage and the magnetization. 
The physical phenomena driven by the spin-dependent Seebeck effect and the anomalous Hall effect, 
such as a generation of electric voltage and excitation of spin torque, have been extensively studied \cite{hatami07,hatami09,xiao10,adachi10,adachi11,taniguchi15}. 
On the other hand, such phenomena by the anomalous Nernst effect have not been quantitatively discussed enough. 
For example, it is still unknown how much temperature gradient is necessary to realize observable phenomena, 
such as a power generation and spin torque switching, by the anomalous Nernst effect, 
which are already reported in the other systems. 
Accordingly, it is fundamentally interesting to 
investigate spin-dependent transport theory driven by the anomalous Nernst effect. 


In this paper, we develop a theory of the spin transport 
in a ferromagnetic/nonmagnetic metallic multilayer by the anomalous Nernst effect. 
Using the phenomenological expression of the electric current carried by th spin-$\nu$ ($\nu=\uparrow,\downarrow$ or $\pm$) electrons 
with the spin-dependent transport coefficients 
and appropriate boundary conditions, 
we evaluate the magnitudes of electric voltage via the inverse spin Hall effect and spin torque 
generated by the anomalous Nernst effect in a ferromagnetic/nonmagnetic multilayer. 
The electric voltage is on the order of $0.1$ $\mu$V for currently available parameter values found in experiments. 
This value is too small for practical application, 
but will be useful in evaluating the spin polarization of the anomalous Nernst coefficient. 
The direction of the spin torque can be controlled by changing the magnetization direction 
which becomes an advantage over the spin torque excitation by the spin Hall effect, where the torque direction is geometrically restricted. 
The magnitude of the temperature gradient for switching ($10^{6}$ K/mm) is, however, quite larger than the experimentally available value. 
In experiments, the conventional Seebeck effect and the transverse spin Seebeck effect will also contribute to the voltage generationa and spin torque effect. 
The procedures to separate thse contributions from that of the anomalous Nernst effect are also discussed. 




\section{Spin transport theory}
\label{sec:Spin transport theory}


\subsection{Definition of spin-dependent current}
\label{sec:Definition of spin-dependent current}

The spin-$\nu$ electron ($\nu=\uparrow,\downarrow$ or $\pm$) in a ferromagnet 
in the presence of the temperature gradient carries the electric current density given by 
\begin{equation}
  \mathbf{J}_{{\rm c},\nu}
  =
  \frac{\sigma_{\nu}}{e}
  \bm{\nabla}
  \bar{\mu}_{\nu}
  +
  \sigma_{\nu}
  S_{\nu}
  \bm{\nabla} 
  T 
  +
  \sigma_{\nu}
  N_{\nu} 
  \mathbf{m}
  \times
  \bm{\nabla} T,
  \label{eq:current_def}
\end{equation}
where $\sigma_{\nu}$, $S_{\nu}$, and $N_{\nu}$ are 
the conductivity, the Seebeck coefficient, and the anomalous Nernst coefficient of the spin-$\nu$ electron, respectively. 
Although we mainly focus on the spin current generated by the anomalous Nernst effect, 
terms related to the spin-dependent Seebeck effect are taken into account as much as possible, for generality. 
The electron's charge is $-e$. 
The unit vector pointing in the direction of the magnetization is denoted as $\mathbf{m}$, 
where we assume that $\mathbf{m}$ is uniform in a ferromagnet. 
The electrochemical potential of the spin-$\nu$ electron is $\bar{\mu}_{\nu}$. 
The spin polarization of the conductivity is defined as \cite{valet93}
\begin{equation}
  \beta
  =
  \frac{\sigma_{\uparrow}-\sigma_{\downarrow}}{\sigma_{\uparrow}+\sigma_{\downarrow}}.
  \label{eq:spin_polarization_sigma}
\end{equation}


To define the electric and spin currents from Eq. (\ref{eq:current_def}), 
it is convenient to introduce the Seebeck coefficient $\mathscr{S}$, the anomalous Nernst coefficient $\mathscr{N}$, 
and their spin polarizations, $p_{\rm S}$ and $p_{\rm N}$, as follows: 
\begin{equation}
  \mathscr{S}
  =
  \frac{\sigma_{\uparrow}S_{\uparrow}+\sigma_{\downarrow}S_{\downarrow}}{\sigma},
  \label{eq:S_2}
\end{equation}
\begin{equation}
  \mathscr{N}
  =
  \frac{\sigma_{\uparrow}N_{\uparrow}+\sigma_{\downarrow}N_{\downarrow}}{\sigma},
  \label{eq:N_2}
\end{equation}
\begin{equation}
  p_{\rm S}
  =
  \frac{\sigma_{\uparrow} S_{\uparrow} - \sigma_{\downarrow} S_{\downarrow}}{\sigma_{\uparrow} S_{\uparrow} + \sigma_{\downarrow} S_{\downarrow}},
  \label{eq:polarization_S_2}
\end{equation}
\begin{equation}
  p_{\rm N}
  =
  \frac{\sigma_{\uparrow} N_{\uparrow} - \sigma_{\downarrow} N_{\downarrow}}{\sigma_{\uparrow} N_{\uparrow} + \sigma_{\downarrow} N_{\downarrow}},
  \label{eq:polarization_N_2}
\end{equation}
where $\sigma=\sigma_{\uparrow}+\sigma_{\downarrow}$. 
Then, Eq. (\ref{eq:current_def}) can be rewritten as 
\begin{equation}
\begin{split}
  \mathbf{J}_{{\rm c},\nu}
  =&
  \frac{(1+\nu \beta)}{2}
  \frac{\sigma}{e}
  \bm{\nabla}
  \bar{\mu}_{\nu}
  +
  \frac{(1+\nu p_{\rm S})}{2}
  \sigma
  \mathscr{S}
  \bm{\nabla} 
  T 
\\
  &+
  \frac{(1+\nu p_{\rm N})}{2}
  \sigma
  \mathscr{N}
  \mathbf{m}
  \times
  \bm{\nabla} T.
  \label{eq:current_def_sub}
\end{split}
\end{equation}
The total electric current density, $\mathbf{J}_{\rm c}=\mathbf{J}_{{\rm c},\uparrow}+\mathbf{J}_{{\rm c},\downarrow}$, 
and the spin current density, $\mathbf{J}_{\rm s}=-[\hbar/(2e)](\mathbf{J}_{{\rm c},\uparrow}-\mathbf{J}_{{\rm c},\downarrow})$, are given by 
\begin{equation}
\begin{split}
  \mathbf{J}_{\rm c}
  =&
  \frac{\sigma}{e}
  \bm{\nabla}
  \bar{\mu}
  +
  \frac{\beta \sigma}{e}
  \bm{\nabla}
  \delta
  \mu
\\
  &+
  \sigma
  \mathscr{S}
  \bm{\nabla} T 
  +
  \sigma
  \mathscr{N}
  \mathbf{m}
  \times
  \bm{\nabla} T,
  \label{eq:electric_current_def}
\end{split}
\end{equation}
\begin{equation}
\begin{split}
  \mathbf{J}_{\rm s}
  =&
  -\frac{\hbar \beta \sigma}{2e^{2}}
  \bm{\nabla}
  \bar{\mu}
  -
  \frac{\hbar \sigma}{2e^{2}}
  \bm{\nabla}
  \delta
  \mu
\\
  &-
  \frac{\hbar p_{\rm S} \sigma}{2e}
  \mathscr{S}
  \bm{\nabla} 
  T 
  -
  \frac{\hbar p_{\rm N} \sigma}{2e}
  \mathscr{N}
  \mathbf{m}
  \times
  \bm{\nabla} T.
  \label{eq:spin_current_def}
\end{split}
\end{equation}
Here, we introduce the electrochemical potential $\bar{\mu}$ and the spin accumulation $\delta\mu$ as 
$\bar{\mu}=(\bar{\mu}_{\uparrow}+\bar{\mu}_{\downarrow})/2$ and $\delta\mu=(\bar{\mu}_{\uparrow}-\bar{\mu}_{\downarrow})/2$. 
The anomalous Nernst coefficient has been estimated experimentally from an electric voltage proportional to $\rho |\mathbf{J}_{\rm c}|$, 
where $\rho=1/\sigma$ is the resistivity \cite{sakuraba13,sakuraba16}. 
Therefore, according to Eq. (\ref{eq:electric_current_def}), 
the experimentally estimated anomalous Nernst coefficient corresponds to $\mathscr{N}$. 
The value of $\beta$ is typically between -1 and 1, 
meaning that both the spin-up and spin-down electrons move in same direction in the presence of the electric field. 
On the other hand, we assume that the value $p_{\rm N}$ does not have such restriction, as in the case of the anomalous Hall effect \cite{taniguchi15}. 
The vector notation in Eq. (\ref{eq:spin_current_def}) represents the direction of the electrons flow in the spatial space. 
We assume that the penetration depth of the transverse spin current is sufficiently short \cite{slonczewski96,brataas01,taniguchi08,ghosh12}, 
and therefore, the direction of the spin polarization of the spin current is parallel to the magnetization. 
The spin current is a tensor product between the spin polarization and electrons flow, i.e., $\mathbf{m} \otimes \mathbf{J}_{\rm s}$. 


Equations (\ref{eq:electric_current_def}) and (\ref{eq:spin_current_def}) are applicable to metallic ferromagnetic/nonmagnetic multilayer 
for the investigation of the spin-dependent physical phenomena, 
such as generation of electric power by spin current and excitation of spin torque. 
We will show examples of such theoretical study in the following. 
In this section, we assume an uniform temperature gradient along the $x$-direction, 
i.e., $\bm{\nabla}T=\partial_{x}T \mathbf{e}_{x}$ and $\bm{\nabla}^{2}T=0$. 
Also, we focus on the spin injection in the $z$-direction, 
and assume translation symmetry along the $y$-direction. 
Using the conservation law of the electric current $\bm{\nabla}\cdot\mathbf{J}_{\rm c}=0$, 
and applying the open circuit conditions along the $x$- and $z$-directions, we find from Eqs. (\ref{eq:electric_current_def}) and (\ref{eq:spin_current_def}) that 
\begin{equation}
  \mathbf{e}_{x}
  \cdot
  \mathbf{J}_{\rm s}
  =
  -\frac{\hbar (1-\beta^{2}) \sigma}{2e^{2}}
  \partial_{x}
  \delta
  \mu
  +
  \frac{\hbar  (\beta-p_{\rm S}) \sigma}{2e}
  \mathscr{S}
  \partial_{x} T,
  \label{eq:spin_current_x}
\end{equation}
\begin{equation}
  \mathbf{e}_{z}
  \cdot
  \mathbf{J}_{\rm s}
  =
  -\frac{\hbar(1-\beta^{2})\sigma}{2e^{2}}
  \partial_{z}
  \delta
  \mu
  -
  \frac{\hbar (\beta-p_{\rm N}) \sigma}{2e}
  \mathscr{N}
  m_{y}
  \partial_{x} T.
  \label{eq:spin_current_z}
\end{equation}
Here, the spin current densities are expressed in terms of the gradients of the spin accumulation and temperature. 
The spin accumulation obeys the diffusion equation \cite{valet93}, 
$\bm{\nabla}^{2}\delta\mu = \delta\mu/\ell^{2}$, with the spin diffusion length $\ell$. 
Note that an additional term proportional to $\bm{\nabla}^{2}T$ should be added to the diffusion equation of the spin accumulation \cite{scharf12} 
when the temperature gradient is nonuniform, $\bm{\nabla}^{2}T \neq 0$. 
In this case, the solution of the spin accumulation depends on the temperature profile. 
In this section, we consider the uniform temperature gradient, as assumed in the experiments \cite{uchida10}. 
For generality, however, we derive the diffusion equation for the nonuniform temperature profile in Appendix \ref{sec:AppendixA}. 
The above formulas are valid in the presence of both the spin-dependent Seebeck and anomalous Nernst effects. 
The physical phenomena related to the spin-dependent Seebeck effect has been extensively studied \cite{uchida08,uchida10,slachter10,bauer12}. 
The following part of this paper, on the other hand, focuses on the anomalous Nernst effect. 
The separation of the contribution from the spin-dependent Seebeck \cite{schmid13,kikkawa13} is discussed below.


The anomalous Hall effect and the anisotropic magnetoresistance in ferromagnets also generate spin current. 
In the presence of these effects, terms proportional to $\sigma_{\rm AHE} \mathbf{m} \times \bm{\nabla}\bar{\mu}_{\nu}$ 
and $\sigma_{\rm AMR} (\mathbf{m}\cdot\bm{\nabla}\bar{\mu}_{\nu})\mathbf{m}$ should be added to Eq. (\ref{eq:current_def}), 
where $\sigma_{\rm AHE}$ and $\sigma_{\rm AMR}$ are the conductivities related to the anomalous Hall effect and the anisotropic magnetoresistance. 
As discussed in Ref. \cite{taniguchi15}, in the presence of an external electric field $E_{x}$ along the $x$-direction, 
the dominant term of the electrochemical potential, $\bm{\nabla}\bar{\mu}_{\nu}$, 
is the electric field, $E_{x} \mathbf{e}_{x}$. 
Then, the anomalous Hall effect, for example, generates pure spin current in the $z$-direction when the magnetization points to the $y$-direction. 
In the present system, on the other hand, the external electric field is absent. 
The electrochemical potential varies spatially along the $z$-direction due to the spin diffusion driven by the anomalous Nernst effect. 
Therefore, the gradient $\bm{\nabla}\bar{\mu}_{\nu}$ points to the $z$-direction. 
In this case, the anomalous Hall effect does not induce spin current in the $z$-direction, and thus, can be neglected here. 




\begin{figure}
\centerline{\includegraphics[width=1.0\columnwidth]{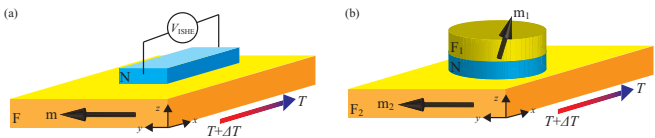}}
\caption{
         (Color online)
         (a) Schematic view of the ferromagnetic(F)/nonmagnetic(N) multilayer. 
             The temperature gradient applied to the ferromagnet along the $x$-direction 
             induces the spin current into the nonmagnet by the anomalous Nernst effect. 
             The injected spin current is converted to an electric voltage $V_{\rm ISHE}$ by the inverse spin Hall effect. 
             The unit vector pointing in the magnetization direction, $\mathbf{m}$, has a finite $y$ component to injects the spin current into the nonmagnet. 
         (b) Schematic view of the ferromagnetic(F${}_{1}$)/nonmagnetic(N)/ferromagnetic(F${}_{2}$) multilayer. 
             The temperature gradient is applied to the F${}_{2}$ layer along the $x$-direction, 
             and induces the spin current into the F${}_{2}$ by the anomalous Nernst effect. 
             The injected spin current excites spin torque on the magnetization of the F${}_{1}$ layer, $\mathbf{m}_{1}$. 
         \vspace{-3ex}}
\label{fig:fig1}
\end{figure}




\subsection{Voltage generation by inverse spin Hall effect}
\label{sec:Voltage generation by inverse spin Hall effect}

A common method to detect spin current is the inverse spin Hall effect, 
where the spin current is converted to an electric voltage. 
In this section, we derive an analytical formula of the electric voltage generated by the anomalous Nernst and inverse spin Hall effects. 


The system we consider is shown in Fig. \ref{fig:fig1}(a), 
where an uniform temperature gradient is applied to a ferromagnet along the $x$-direction, 
while a nonmagnet is placed on the ferromagnet along the $z$-direction. 
Here, we use the subscripts, "F" and "N", to distinguish 
whether the quantities are defined in the ferromagnetic or in the nonmagnetic layer. 
A temperature gradient generates spin current flowing into the nonmagnet by the anomalous Nernst effect, 
and the injected spin current is converted to the electric voltage by the inverse spin Hall effect. 
In the spin-dependent Seebeck experiment \cite{uchida08}, the magnetization points to the $x$-direction, 
and the inverse spin Hall effect generates an electric voltage along the $y$-direction. 
On the other hand, the magnetization $\mathbf{m}$ has a finite $y$ component to inject the spin current into the nonmagnet. 
In this case, the inverse spin Hall effect generates the electric voltage along the $x$-direction. 


Solving the diffusion equation, 
the solution of the spin accumulation in the ferromagnet is given by 
\begin{equation}
\begin{split}
  \delta
  \mu_{\rm F}
  =&
  \frac{4\pi}{2(g_{\rm F}/A) \sinh(d_{\rm F}/\ell_{\rm F})}
\\
  & \times
  \left\{
    \frac{\hbar(\beta-p_{\rm N})}{2e}
    \sigma_{\rm F}
    \mathscr{N}
    m_{y}
    \partial_{x} T 
    \cosh
    \left(
      \frac{z-d_{\rm F}}{\ell_{\rm F}}
    \right)
  \right.
\\
  &
  \left.
    -
    \left[
      J_{\rm s}^{\rm F/N}
      +
      \frac{\hbar (\beta-p_{\rm N})}{2e}
      \sigma_{\rm F}
      \mathscr{N}
      m_{y}
      \partial_{x} T
    \right]
    \cosh
    \left(
      \frac{z}{\ell_{\rm F}}
    \right)
  \right\},
  \label{eq:solution_spin_accumulation_F}
\end{split}
\end{equation}
where $J_{\rm s}^{\rm F/N}$ is the spin current density flowing at the ferromagnetic/nonmagnetic boundary. 
The thicknesses of the ferromagnet and nonmagnet along the $z$-direction are denoted as $d_{\rm F}$ and $d_{\rm N}$, respectively. 
We also introduce $g_{\rm F}/A=h(1-\beta^{2})\sigma_{\rm F}/(2e^{2}\ell_{\rm F})$, 
where $A$ is the cross section area of the ferromagnetic/nonmagnetic interface. 
The typical nonmagnets used in the experiments of the inverse spin Hall effect are, for example, Pt, Ta, and W 
because these heavy metals show large spin Hall angles. 
The spin diffusion length of such a heavy metal nonmagnet is usually short \cite{bass07}, 
and therefore, the diffusion equation of the spin accumulation in the nonmagnet should also be solved. 
Similarly to Eq. (\ref{eq:solution_spin_accumulation_F}), the spin accumulation in the nonmagnet is given by 
\begin{equation}
  \delta
  \mu_{\rm N}
  =
  \frac{4\pi}{2(g_{\rm N}/A)\sinh(d_{\rm N}/\ell_{\rm N})}
  J_{\rm s}^{\rm F/N}
  \cosh
  \left(
    \frac{z-d_{\rm F}-d_{\rm N}}{\ell_{\rm N}}
  \right),
  \label{eq:solution_spin_accumulation_N}
\end{equation}
where $g_{\rm N}/A=h\sigma_{\rm N}/(2e^{2}\ell_{\rm N})$. 
The boundary condition at the ferromagnetic/nonmagnetic interface is \cite{valet93,brataas06} 
\begin{equation}
  J_{\rm s}^{\rm F/N}
  =
  \frac{1}{4\pi A}
  \frac{(1-\gamma^{2})g}{2}
  2 
  \left[
    \delta \mu_{\rm F}(z=d_{\rm F})
    -
    \delta \mu_{\rm N}(z=d_{\rm F})
  \right],
  \label{eq:boundary_condition_ISHE}
\end{equation}
where $g=g^{\uparrow}+g^{\downarrow}$ and $\gamma=(g^{\uparrow}-g^{\downarrow})/(g^{\uparrow}+g^{\downarrow})$ 
are the dimensionless interface conductance and its spin polarization, respectively. 
The conductance $g^{\nu}$ ($\nu=\uparrow,\downarrow$) relates to the interface resistance of the spin-$\nu$ electron as $r_{\nu}=(h/e^{2})A/g^{\nu}$ 
with $h/e^{2} \simeq 25.9$ k$\Omega$. 
From Eq. (\ref{eq:boundary_condition_ISHE}), the spin current density at the interface is 
\begin{equation}
  J_{\rm s}^{\rm F/N}
  =
  -\frac{\hbar(\beta-p_{\rm N}) \tilde{g}}{2e g_{\rm F}}
  \sigma_{\rm F} 
  \mathscr{N}
  m_{y}
  \partial_{x} T 
  \tanh
  \left(
    \frac{d_{\rm F}}{2 \ell_{\rm F}}
  \right),
\end{equation}
where $\tilde{g}$ is defined as 
\begin{equation}
  \frac{1}{\tilde{g}}
  =
  \frac{2}{(1-\gamma^{2})g}
  +
  \frac{1}{g_{\rm F} \tanh(d_{\rm F}/\ell_{\rm F})}
  +
  \frac{1}{g_{\rm N} \tanh(d_{\rm N}/\ell_{\rm N})}.
\end{equation}
Then, we find that the spin current density inside the nonmagnet, $J_{\rm s,N}=-[\hbar \sigma_{\rm N}/(2e^{2})] \partial_{z} \delta\mu_{\rm N}$, is 
\begin{equation}
\begin{split}
  J_{\rm s,N}
  =&
  \frac{\hbar(\beta-p_{\rm N}) \tilde{g}}{2e g_{\rm F}}
  \frac{\tanh[d_{\rm F}/(2 \ell_{\rm F})]}{\sinh(d_{\rm N}/\ell_{\rm N})}
\\
  & \times
  \sigma_{\rm F}
  \mathscr{N}
  m_{y}
  \partial_{x} T 
  \sinh
  \left(
    \frac{z-d_{\rm F}-d_{\rm N}}{\ell_{\rm N}}
  \right).
\end{split}
\end{equation}
The averaged spin current density, $\langle J_{\rm s,N} \rangle = (1/d_{\rm F})\int_{d_{\rm F}}^{d_{\rm F}+d_{\rm N}} J_{\rm s,z} dz$, becomes 
\begin{equation}
\begin{split}
  \left\langle
    J_{\rm s,N}
  \right\rangle
  =&
  -\frac{\hbar (\beta-p_{\rm N}) \tilde{g} \ell_{\rm N}}{2e g_{\rm F} d_{\rm N}}
  \tanh
  \left(
    \frac{d_{\rm F}}{2 \ell_{\rm F}}
  \right)
\\
  & 
  \times
  \tanh
  \left(
    \frac{d_{\rm N}}{2 \ell_{\rm N}}
  \right)
  \sigma_{\rm F}
  \mathscr{N}
  m_{y}
  \partial_{x}T. 
  \label{eq:average_current}
\end{split}
\end{equation}
The spin current is converted to an electric current by the inverse spin Hall effect as 
$\mathbf{J}_{\rm c,ISHE}=\vartheta \mathbf{m} \times \mathbf{J}_{\rm s}/[-\hbar/(2e)]$, 
where $\vartheta$ is the spin Hall angle in the nonmagnet, 
and we assume that the magnetization is assumed to be parallel to the $y$-direction. 
This electric current generates an electric field along the $x$-direction given by 
$E_{\rm ISHE}=(\vartheta/\sigma_{\rm N}) m_{y} \langle J_{\rm s,N} \rangle/[-\hbar/(2e)]$. 
Then, we find that an electric voltage generated through the inverse spin Hall effect, $V_{\rm ISHE}=E_{\rm ISHE}L_{N}$, is given by 
\begin{equation}
\begin{split}
  V_{\rm ISHE}
  =&
  \frac{\vartheta(\beta-p_{\rm N}) \tilde{g} \ell_{\rm N} L_{\rm N}}{\sigma_{\rm N} g_{\rm F} d_{\rm N}}
  \tanh
  \left(
    \frac{d_{\rm F}}{2\ell_{\rm F}}
  \right)
\\
  &
  \times
  \tanh
  \left(
    \frac{d_{\rm N}}{2\ell_{\rm N}}
  \right)
  \sigma_{\rm F}
  \mathscr{N}
  m_{y}^{2}
  \partial_{x}T,
  \label{eq:V_ISHE}
\end{split}
\end{equation}
where $L_{\rm N}$ is the length of the nonmagnet along the $x$-direction. 
We should emphasize here that the large aspect ratio of the nonmagnet, $L_{\rm N}/d_{\rm N}$, in Eq. (\ref{eq:V_ISHE}) enables to generate an observable voltage, 
although the spin current magnitude is small. 
Therefore, it will be possible to measure a small voltage generated by a reasonable temperature gradient by optimizing the experimental geometry. 
Such approach cannot be used to the spin torque switching problem, 
and therefore, it is difficult to excite the magnetization dynamics by the spin torque driven by the temperature gradient, as will be discussed in the next section. 




\begin{figure}
\centerline{\includegraphics[width=1.0\columnwidth]{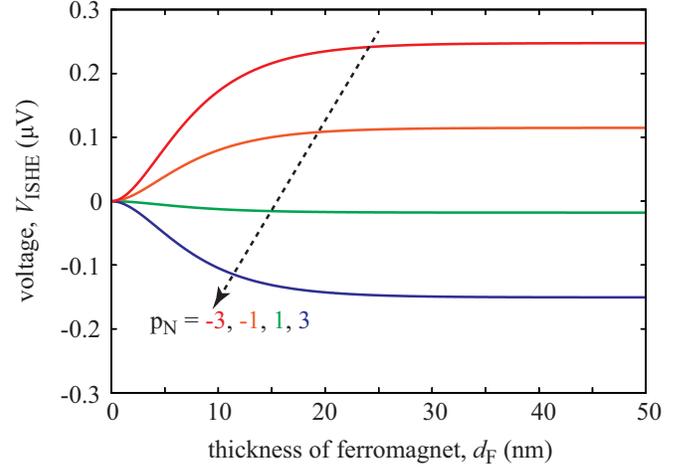}}
\caption{
         (Color online)
         Dependence of the electric voltage $V_{\rm ISHE}$ on the thickness of the ferromagnet, $d_{\rm F}$, 
         where $\mathscr{N}=1.0$ $\mu$V/K and $\partial_{x}T=30$ K/mm. 
         The value of the spin polarization of the anomalous Nernst coefficient $p_{\rm N}$ varies as 3, -1, 1, and 3. 
         \vspace{-3ex}}
\label{fig:fig2}
\end{figure}



A typical value of the electric voltage obtained in the spin-dependent Seebeck experiments is on the order of 1 $\mu$V \cite{uchida08}. 
Now let us estimate the temperature gradient to generate the electric voltage of 1 $\mu$V from the anomalous Nernst effect. 
The experimental value of $\mathscr{N}$ is on the order of 1 $\mu$V/K \cite{sakuraba13,sakuraba16}, 
which is one order of magnitude smaller than the Seebeck coefficient. 
Considering NiFe/Pt multilayer as an example, 
the resistivity $\rho_{\rm F}=1/\sigma_{\rm F}$, the spin polarization $\beta$, and the spin diffusion length $\ell_{\rm F}$ of the ferromagnet are 
241 $\Omega$nm, 0.73, and 5.5 nm, respectively \cite{uchida09,bass07}, 
We assume that $p_{\rm N}=3$ for NiFe \cite{uchida09}. 
The magnetization points to the $y$-direction, $m_{y}=1$. 
On the other hand, for Pt, $\rho_{\rm N}=1/\sigma_{\rm N}$, $\ell_{\rm N}$, and $\vartheta$ are 
397 $\Omega$nm, 2 nm, and 0.01, respectively \cite{isasa15}. 
The thicknesses are assumed to be $d_{\rm F}=10$ nm and $d_{\rm N}=10$ nm, 
while $L_{\rm N}=1$ mm, which are similar to the values used in the spin-dependent Seebeck experiments. 
The value of the interface resistance is derived from typical metallic ferromagnetic/nonmagnetic multilayer \cite{brataas06}, 
$(1-\gamma^{2})g/(2A)=24$ nm${}^{-2}$. 
Then, we find that the temperature gradient satisfying $V_{\rm ISHE}=1$ $\mu$V is on the order of 
$|\partial_{x}T|=2.9 \times 10^{2}$ K/mm. 
This value is one order of magnitude larger than the temperature gradient used in spin-dependent Seebeck experiments, 10 K/mm. 
The large value of the temperature gradient is due to the small value of the anomalous Nernst coefficient, $\mathscr{N} \sim 0.1 \mathscr{S}$. 
In other words, the electric voltage generated by the anomalous Nernst effect for currently available values of the parameters 
is on the order of 0.1 $\mu$V when a conventional temperature gradient is applied. 


Although the generated electric voltage is small, 
the measurement of the electric voltage might provide important information of the spin-dependent transport properties caused by the anomalous Nernst effect. 
Figure \ref{fig:fig2} shows the dependence of the electric voltage, Eq. (\ref{eq:V_ISHE}), on the thickness of the ferromagnet $d_{\rm F}$. 
The values of the parameters are those mentioned above, while the temperature gradient is 30 K/mm. 
The magnitude of $V_{\rm ISHE}$ increases from zero, and saturates for $d_{\rm F} \gg \ell_{\rm F}$. 
Note that the saturated value of $V_{\rm ISHE}$ is proportional to $\beta-p_{\rm N}$. 
Therefore, the measurement of $V_{\rm ISHE}$ will be useful to evaluate the magnitude and the sign of the spin polarization of the anomalous Nernst effect. 



\subsection{Excitation of spin torque}
\label{sec:Excitation of spin torque}

In this section, we study the excitation of spin torque by the anomalous Nernst effect. 
The system we consider is shown in Fig. \ref{fig:fig1}(b), 
where a nonmagnet, N, is sandwiched by two ferromagnets, F${}_{1}$ and F${}_{2}$ layers. 
The temperature gradient is applied to the F${}_{2}$ layer. 
The F${}_{2}$ layer injects spin current into the F${}_{1}$ layer, 
and excites spin torque on the magnetization of the F${}_{1}$ layer, $\mathbf{m}_{1}$. 
In the following, we use the subscripts "$k$" or "F${}_{k}$" ($k=1,2$) to distinguish 
whether the quantities are related to the F${}_{k}$ layer or to the F${}_{k}$/N interface. 


The spin torque is excited by the injection of the spin current through the interface. 
Similar to the previous section, the spin current at the ferromagnetic/nonmagnetic interface 
relates to the spin accumulations in each layer. 
Note that the boundary condition of the spin current and spin accumulation used in the previous section, Eq. (\ref{eq:boundary_condition_ISHE}), 
is valid when the alignment of the magnetizations in the multilayer is collinear. 
On the other hand, a noncollinear alignment of the magnetizations is necessary to excite spin torque \cite{slonczewski96,berger96}. 
In a ferromagnetic multilayer having noncollinear alignment of the magnetizations, 
Eq. (\ref{eq:boundary_condition_ISHE}) is extended to 
\begin{equation}
\begin{split}
  \mathbf{I}_{\rm s}^{{\rm F}_{k} \to {\rm N}}
  =
  \frac{1}{4\pi}
  &
  \left[
    \frac{(1-\gamma_{k}^{2})g_{k}}{2}
    \mathbf{m}_{k}
    \cdot
    \left(
      \bm{\mu}_{{\rm F}_{k}}
      -
      \bm{\mu}_{\rm N}
    \right)
    \mathbf{m}_{k}
  \right.
\\
  &-
  \left.
    g_{{\rm r}({\rm F}_{k})}
    \mathbf{m}_{k}
    \times
    \left(
      \bm{\mu}_{{\rm F}_{k}}
      \times
      \mathbf{m}_{k}
    \right)
    -
    g_{{\rm i}({\rm F}_{k})}
    \bm{\mu}_{\rm N}
    \times
    \mathbf{m}_{k}
  \right],
  \label{eq:spin_current_FN}
\end{split}
\end{equation}
where $g_{\rm r}$ and $g_{\rm i}$ are the real and imaginary parts of the mixing conductance \cite{brataas06}. 
The vector notation in Eq. (\ref{eq:spin_current_FN}) represents the spin polarization of spin current flowing in the $z$-direction. 
Note that $\bm{\mu}_{\rm F}$ is defined as $\mathbf{m}\cdot\bm{\mu}_{\rm F}=\bar{\mu}_{\uparrow}-\bar{\mu}_{\downarrow}$, according to Ref. \cite{brataas06}, 
while $\delta\mu$ in the previous section is $\delta\mu=(\bar{\mu}_{\uparrow}-\bar{\mu}_{\downarrow})/2$. 
Substituting the solution of the spin accumulation in the F${}_{2}$ layer, which is similar to Eq. (\ref{eq:solution_spin_accumulation_F}), into Eq. (\ref{eq:spin_current_FN}), 
the spin current at the F${}_{2}$/N interface is 
\begin{equation}
\begin{split}
  \mathbf{I}_{\rm s}^{\rm F_{2} \to N}
  =&
  -\frac{\hbar g_{\rm F_{2}}^{*}}{2e g_{\rm F_{2}}}
  \tanh
  \left(
    \frac{d_{2}}{2 \ell_{2}}
  \right)
  (\beta-p_{\rm N})
  \sigma_{\rm F_{2}}
  \mathscr{N}
  m_{2y}
  A
  \partial_{x}T
  \mathbf{m}_{2}
\\
  &-
  \frac{1}{4\pi}
  \left[
    g_{\rm F_{2}}^{*}
    \left(
      \mathbf{m}_{2}
      \cdot
      \bm{\mu}_{\rm N}
    \right)
    \mathbf{m}_{2}
    +
    g_{\rm r(F_{2})}
    \mathbf{m}_{2}
    \times
    \left(
      \bm{\mu}_{\rm N}
      \times
      \mathbf{m}_{2}
    \right)
  \right.
\\
  &
  \left.
    +
    g_{\rm i(F_{2})}
    \bm{\mu}_{\rm N}
    \times
    \mathbf{m}_{2}
  \right],
  \label{eq:spin_current_F2N}
\end{split}
\end{equation}
where $g_{{\rm F}_{k}}/A=h(1-\beta^{2})\sigma_{{\rm F}_{k}}/(2e^{2}\ell_{k})$ was introduced in the previous section, 
while $g_{{\rm F}_{k}}^{*}$ is defined as 
\begin{equation}
  \frac{1}{g_{{\rm F}_{k}}^{*}}
  =
  \frac{2}{(1-\gamma_{{\rm F}_{k}}^{2})g_{k}}
  +
  \frac{1}{g_{{\rm F}_{k}}\tanh(d_{k}/\ell_{k})}.
\end{equation}
On the other hand, the spin current at the F${}_{1}$/N interface is given by 
\begin{equation}
\begin{split}
  \mathbf{I}_{\rm s}^{\rm F_{1} \to N}
  =
  -\frac{1}{4\pi}
  &
  \left[
    g_{\rm F_{1}}^{*}
    \left(
      \mathbf{m}_{1}
      \cdot
      \bm{\mu}_{\rm N}
    \right)
    +
    g_{\rm r(F_{1})}
    \mathbf{m}_{1}
    \times
    \left(
      \bm{\mu}_{\rm N}
      \times
      \mathbf{m}_{1}
    \right)
  \right.
\\
  &
  \left.
    +
    g_{\rm i(F_{1})}
    \bm{\mu}_{\rm N}
    \times
    \mathbf{m}_{1}
  \right].
  \label{eq:spin_current_F1N}
\end{split}
\end{equation}


A nonmagnetic metal having a long spin diffusion length, such as Cu \cite{bass07}, should be inserted between the F${}_{1}$ and F${}_{2}$ layers 
to avoid the relaxation of the spin polarization of the spin current. 
Thus, we assume that the spin current in the nonmagnet is conserved, 
i.e., $\mathbf{I}_{\rm s}^{\rm F_{1} \to N}+\mathbf{I}_{\rm s}^{\rm F_{2} \to N}=\bm{0}$. 
The spin accumulation in the nonmagnet is determined by this condition. 
In general, the spin accumulation in the nonmagnet can be expressed as 
\begin{equation}
  \bm{\mu}_{\rm N}
  =
  a
  \mathbf{m}_{1}
  +
  b
  \mathbf{m}_{1}
  \times
  \mathbf{m}_{2}
  +
  c
  \mathbf{m}_{1}
  \times
  \left(
    \mathbf{m}_{2}
    \times
    \mathbf{m}_{1}
  \right). 
  \label{eq:spin_accumulation_expand}
\end{equation}
The exact solutions of the coefficients, $a$, $b$, and $c$, are summarized in Appendix \ref{sec:AppendixB}. 
The spin torque acting on the magnetization of the F${}_{1}$ layer relates to the spin accumulation $\bm{\mu}_{\rm N}$ via 
\begin{equation}
\begin{split}
  \mathbf{T}
  &=
  \frac{\gamma_{0}}{M_{1}V_{1}}
  \mathbf{m}_{1}
  \times
  \left(
    \mathbf{I}_{\rm s}^{\rm F_{1} \to N}
    \times
    \mathbf{m}_{1}
  \right)
\\
  &=
  -\frac{\gamma_{0}\hbar}{4\pi M_{1}V_{1}}
  \left[
    g_{\rm r(F_{1})}
    \mathbf{m}_{1}
    \times
    \left(
      \bm{\mu}_{\rm N}
      \times
      \mathbf{m}_{1}
    \right)
    +
    g_{\rm i(F_{1})}
    \bm{\mu}_{\rm N}
    \times
    \mathbf{m}_{1}
  \right],
\end{split}
\end{equation}
where $\gamma_{0}$, $M_{1}$, and $V_{1}=Ad_{1}$ are the gyromagnetic ratio, saturation magnetization, and volume of the F${}_{1}$ layer, respectively. 
Assuming a negligibly small value for the imaginary part of the mixing conductance $g_{\rm i}$ \cite{brataas06}, 
we find that the spin torque formula is given by 
\begin{equation}
\begin{split}
  \mathbf{T}
  =
  -&\frac{\gamma_{0}\hbar}{2eM_{1}d_{1}}
  \frac{g_{\rm r(F_{1})}g_{\rm F_{2}}^{*} \tanh[d_{2}/(2 \ell_{2})]}{g_{\rm F_{2}}[g_{\rm r(F_{1})}+g_{\rm F_{2}}^{*}]}
\\
  & \times
  (\beta - p_{\rm N})
  \sigma_{\rm F_{2}}
  \mathscr{N}
  m_{2y}
  \partial_{x}T
  \frac{\mathbf{m}_{1} \times (\mathbf{m}_{2} \times \mathbf{m}_{1})}{1-\lambda_{1}\lambda_{2}(\mathbf{m}_{1}\cdot\mathbf{m}_{2})^{2}}. 
  \label{eq:STT}
\end{split}
\end{equation}
The parameter $\lambda_{k}$ is defined as ($(k,k^{\prime})=(1,2)$ or $(2,1)$)
\begin{equation}
  \lambda_{k}
  =
  \frac{g_{{\rm r}({\rm F}_{k})} - g_{{\rm F}_{k}}^{*}}
    {g_{{\rm r}({\rm F}_{k^{\prime}})} + g_{{\rm F}_{k}}^{*}}. 
\end{equation}
Similar to the spin torque excited by an electric current in a CPP geometry, 
the direction of the spin torque can be controlled by changing the magnetization direction of the F${}_{2}$ layer, $\mathbf{m}_{2}$. 
This is an important difference compared with the spin torque excited by the spin Hall effect \cite{liu12}, 
where the direction of the spin torque is determined geometrically. 
The dependence of the spin torque magnitude on the relative angle of the magnetizations, $\cos^{-1}\mathbf{m}_{1}\cdot\mathbf{m}_{2}$, is, 
on the other hand, different from the spin torque in a CPP geometry, which is described as $\mathbf{m}_{1} \times (\mathbf{m}_{2} \times \mathbf{m}_{1})/(1 + \lambda \mathbf{m}_{1}\cdot\mathbf{m}_{2})$. 
Equation (\ref{eq:STT}) is similar to the spin torque formula excited by the anomalous Hall effect \cite{taniguchi15}. 
Note also that the spin torque driven by the spin-dependent Seebeck effect is studied in Refs. \cite{hatami07,hatami09,jia11}. 


Let us evaluate the temperature gradient necessary for switching the magnetization in the F${}_{1}$ layer. 
The magnetization dynamics in the F${}_{1}$ layer is described by the Landau-Lifshitz-Gilbert (LLG) equation, 
\begin{equation}
  \frac{d \mathbf{m}_{1}}{dt}
  =
  -\gamma_{0}
  \mathbf{m}_{1}
  \times
  \mathbf{H}
  +
  \mathbf{T}
  +
  \alpha
  \mathbf{m}_{1}
  \times
  \frac{d \mathbf{m}_{1}}{dt}.
  \label{eq:LLG}
\end{equation}
The spin torque, $\mathbf{T}$, is given by Eq. (\ref{eq:STT}). 
The Gilbert damping constant $\alpha$ is assumed to be small, 
and therefore, the higher order terms of $\alpha$, 
as well as the products of $\alpha$ and spin torque, are neglected. 
The magnetic field, $\mathbf{H}=H_{\rm K}m_{1z}\mathbf{e}_{z}$, consists of perpendicular anisotropy field $H_{\rm K}$. 
Linearizing the LLG equation around the equilibrium $\mathbf{m}_{1}=+\mathbf{e}_{z}$, 
we find that the temperature gradient necessary to destabilize the magnetization of the F${}_{1}$ layer is given by 
\begin{equation}
\begin{split}
  \partial_{x}T 
  =&
  \frac{2\alpha eM_{1}d_{1}}{\hbar (\beta-p_{\rm N}) \sigma_{\rm F_{2}} \mathscr{N} \tanh[d_{2}/(2 \ell_{2})]}
\\
  &
  \times
  \frac{(1-\lambda_{1}\lambda_{2}m_{2z}^{2})^{2}g_{\rm F_{2}}[g_{\rm r(F_{1})}+g_{\rm F_{2}}^{*}]}{(1-\lambda_{1}\lambda_{2})m_{2y}m_{2z}g_{\rm r(F_{1})}g_{\rm F_{2}}^{*}}
  H_{\rm K}.
  \label{eq:critical_temperature}
\end{split}
\end{equation}
The critical temperature gradient becomes infinite when $m_{2y}=0$ because 
the anomalous Nernst effect does not produce spin current in the $z$-direction. 
The critical temperature gradient becomes also infinite when $m_{2z}=0$ 
since the work done by spin torque during a precession around the equilibrium state is zero. 




\begin{figure}
\centerline{\includegraphics[width=1.0\columnwidth]{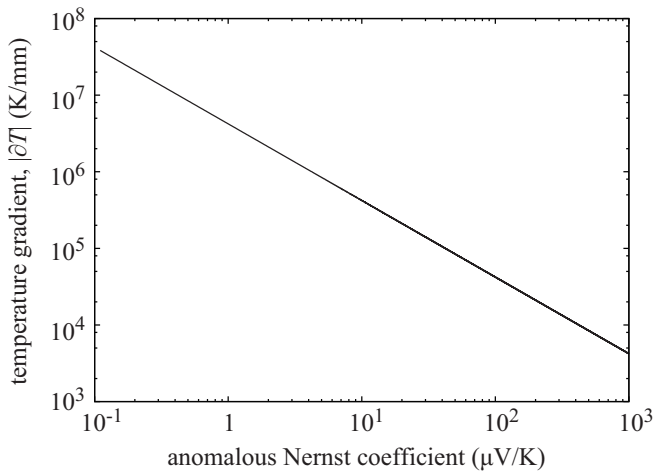}}
\caption{
         The dependence of the magnitude of the temperature gradient, $|\partial_{x}T|$, for spin torque switching on the anomalous Nernst coefficient $\mathscr{N}$. 
         \vspace{-3ex}}
\label{fig:fig3}
\end{figure}



We estimate the magnitude of Eq. (\ref{eq:critical_temperature}) for typical ferromagnet used in the experiments, 
where $M=1000$ emu/c.c., $H_{\rm K}=500$ Oe, and $\alpha=0.005$, respectively \cite{sakuraba13,tsunegi14}. 
The thicknesses of the F${}_{1}$ and F${}_{2}$ layers are assumed to be 2 and 10 nm, respectively. 
The mixing conductance is $g_{\rm r}/A=25$ nm${}^{-2}$ \cite{brataas06}. 
The other parameters, $\rho$, $\ell$, $\beta$, and $(1-\gamma^{2})g/A$ are the same with those used in the previous section. 
The magnetization direction of the F${}_{2}$ layer is set to be $(m_{x},m_{y},m_{z})=(0,\sin 45^{\circ},\cos 45^{\circ})$. 
Figure \ref{fig:fig3} shows the dependence of the temperature gradient $|\partial_{x}T|$, Eq. (\ref{eq:critical_temperature}), 
on the anomalous Nernst coefficient, $\mathscr{N}$. 
The temperature gradient for spin torque switching with a typical value of the anomalous Nernst coefficient, $\mathscr{N}=1 \sim 10$ $\mu$/V, 
is on the order of $10^{5}-10^{6}$ K/mm, 
which is quite larger than the temperature gradient used in current experiments 
for both the spin-dependent Seebeck and anomalous Nernst effects \cite{uchida08,uchida09,sakuraba13,sakuraba16}. 
Therefore, the experimental observation of the spin torque switching by applying a temperature gradient is current difficult. 
An alternative method to produce temperature gradient and/or 
significant improvement in developing suitable material will be necessary to observe the magnetization dynamics excited by the anomalous Nernst effect. 



\subsection{Separation of Seebeck and transverse spin Seebeck effects}
\label{sec:Separation of Seebeck and transverse spin Seebeck effects}

The above calculations have neglected the Seebeck effect and the transverse spin Seebeck effect, 
while these effects are usually much larger than the anomalous Nernst effect in the conventional metals \cite{uchida08,uchida10,slachter10,bauer12,sakuraba13,sakuraba16}. 
In this section, let us discuss the methods to separate the contributions from these effects from the above results. 

We first discuss the contribution from the Seebeck effect to the generation of the electric voltage. 
For example, when the temperature gradient is also applied to the nonmagnet in the geometry of Fig. \ref{fig:fig1}(a), 
the Seebeck effect in the nonmagnet generates an electric voltage along the $x$-direction given by 
$V_{\rm S}=S_{\rm N}L_{N}\partial_{x}T$, where $S_{\rm N}$ is the Seebeck coefficient of the nonmagnet. 
Using the values used in Fig. \ref{fig:fig2}, i.e., $\partial_{x}T=30$ K/mm and $L_{N}=1$ mm, 
and assuming that $S_{\rm N} \simeq 10 \times \mathscr{N} \sim 10$ $\mu$V/K \cite{sakuraba13,sakuraba16}, 
we find that $V_{\rm S}$ is on the order of $10^{2}$ $\mu$V. 
This value is much higher than the voltage generated through the inverse spin Hall effect. 
The contribution from the Seebeck effect, however, is separated by measuring the dependence of the voltage on the thickness of the nonmagnet, $d_{\rm N}$, 
because the contribution from the Seebeck effect is independent of the thickness, 
while that through the inverse spin Hall effect decreases with decreasing the thickness, 
and becomes zero in the zero thickness limit, as can be seen in Eq. (\ref{eq:V_ISHE}). 
In other words, the contribution from the Seebeck effect is evaluated 
from the total voltage $V_{\rm total}$ measured in the geometry of Fig. \ref{fig:fig1}(a) 
as $V_{\rm S}=\lim_{d_{\rm F},d_{\rm N} \to 0}V_{\rm total}$. 

We also discuss the effect of the transverse spin Seebeck effect on the voltage measurement. 
The spin-dependent Seebeck effect creates the spin accumulation in the ferromagnet, 
which is given by 
\begin{equation}
  \delta
  \mu
  =
  \frac{(\beta-p_{\rm S}) e \ell_{\rm F} \mathscr{S}\partial_{x}T}{(1-\beta^{2}) \sinh(L_{\rm F}/\ell_{\rm F})}
  \left[
    \cosh
    \left(
      \frac{x}{\ell_{\rm F}}
    \right)
    -
    \cosh
    \left(
      \frac{x-L_{\rm F}}{\ell_{\rm F}}
    \right)
  \right],
  \label{eq:spin_accumulation_Seebeck}
\end{equation}
where we apply the open circuit condition along the $x$-direction. 
The length of the ferromagnet along the $x$-direction is denoted as $L_{\rm F}$. 
Note that the sign of Eq. (\ref{eq:spin_accumulation_Seebeck}) is changed with respect to the center of the ferromagnet, 
as confirmed in the experiment \cite{uchida08}. 
The spin accumulation given by Eq. (\ref{eq:spin_accumulation_Seebeck}) generates spin current flowing along the $z$-direction through the interface, 
and produces the electric field $E_{\rm ISHE}^{\rm TSSE}$ along the $x$-direction by the inverse spin Hall effect, 
which is proportional to $\delta \mu(x)$ \cite{uchida10s}. 
The voltage generated by the transverse spin Seebeck effect is given by 
$\int E_{\rm ISHE}^{\rm TSSE} dx \propto \int \delta \mu(x) dx$, where the integral range is over the length of the nonmagnet. 
When the center of the nonmagnet along the $x$-direction is same with that of the ferromagnet, 
the integral becomes zero, indicating that the transverse spin Seebeck effect does not generates net voltage along the $x$-direction. 
This is because the electrons having the spin parallel and antiparallel to the magnetization of the ferromagnet 
equally contributes to the voltage with the opposite sign, and cancel each other. 

Alternative method to separate the contribution from the transverse spin Seebeck effect is to measure the voltages 
by changing the magnetization direction from $\mathbf{m}=+\mathbf{e}_{y}$ to $\mathbf{m}=-\mathbf{e}_{y}$. 
In this case, the spin current originated from the spin Seebeck effect changes the direction of the spin polarization, 
resulting in the sign change in the voltage through the inverse spin Hall effect. 
On the other hand, the voltage originated from the anomalous Nernst effect does not change the sign 
because the direction of the electrons flow simultaneously changes its direction. 
This difference can be seen from Eq. (\ref{eq:spin_current_def}). 
The spin current originated from the spin Seebeck effect is the vector product 
between the spin polarization ($\parallel \mathbf{m}$) and the current density, i.e., 
$\propto \mathbf{m} \otimes \mathscr{S}\bm{\nabla}T$, 
and therefore, changes the sign by the reversal of the magnetization direction from $\mathbf{m}=+\mathbf{e}_{y}$ to $\mathbf{m}=-\mathbf{e}_{y}$. 
On the other hand, the spin current originated from the anomalous Nernst effect is proportional to $\mathbf{m} \otimes \mathscr{N} \mathbf{m} \times \bm{\nabla}T$, 
which does not change the sign by the magnetization reversal. 

In summary, the electric voltage measured in the geometry in Fig. \ref{fig:fig1}(a) consists of three contributions, 
the voltages generated through the inverse spin Hall effect via the anomalous Nernst effect $V_{\rm ISHE}^{\rm ANE}$ 
and the transverse spin Seebeck effect $V_{\rm ISHE}^{\rm TSSE}$, 
and the Seebeck effect $V_{\rm S}$. 
The total voltage is $V_{\rm total}=V_{\rm S}+V_{\rm ISHE}^{\rm TSSE}+V_{\rm ISHE}^{\rm ANE}$. 
The contribution from the Seebeck effect is separated by subtracting the voltage in the zero thickness limit, 
$\lim_{d_{\rm N}, d_{\rm F} \to 0}V_{\rm total}=V_{\rm S}$. 
The transverse spin Seebeck effect does not contribute to the voltage when the centers of the nonmagnet and the ferromagnet are same. 
Even if these centers locate at different positions, 
the contribution from the transverse spin Seebeck effect is separated by comparing the voltages for $\mathbf{m}=\pm\mathbf{e}_{y}$, 
i.e., $V_{\rm total}(\mathbf{m}=+\mathbf{e}_{y})-V_{\rm total}(\mathbf{m}=-\mathbf{e}_{y})=2 V_{\rm ISHE}^{\rm TSSE}(\mathbf{m}=+\mathbf{e}_{y})$. 

Next, we consider the separation of the contribution from the transverse spin Seebeck effect to the spin torque switching. 
Similar to the above discussion, the transverse spin Seebeck effect does not contribute to the switching 
when the nonmagnet locates on the center of the F${}_{2}$ layer, 
where the spin accumulation generated by the spin-dependent Seebeck effect, Eq. (\ref{eq:spin_accumulation_Seebeck}), is zero. 
On the other hand, when the nonmagnet locates on different position, the transverse spin Seebeck effect also excites spin torque on the magnetization in the F${}_{1}$ layer. 
The temperature gradient for the magnetization switching is, in general, sum of 
the contributions from the anomalous Nernst effect $\partial_{x}T^{\rm ANE}$ given by Eq. (\ref{eq:critical_temperature}) and 
the transverse spin Seebeck effect $\partial_{x}T^{\rm TSSE}$. 
We notice that the sign of the temperature gradient for the magnetization switching by the anomalous Nernst effect changes its sign 
by changing the magnetization direction from $\mathbf{m}_{2}=(0,m_{2y},m_{2z})$ to $\mathbf{m}_{2}=(0,-m_{2y},m_{2z})$. 
This is because such change of the magnetization direction changes the flow direction of the electrons having the spin parallel and antiparallel to $\mathbf{m}_{2}$, 
and thus, the $z$ component of the spin polarization injected into the F${}_{1}$ layer also changes its sign. 
On the other hand, the $z$ component of the spin polarization in the spin current originated from the transverse spin Seebeck effect does not change its sign 
by such change of the magnetization direction. 
Therefore, the contribution from the anomalous Nernst effect to the magnetization switching is evaluated by comparing the temperature gradients of the switching 
for $\mathbf{m}_{2}=(0,m_{2y},m_{2z})$ and $\mathbf{m}_{2}=(0,-m_{2y},m_{2z})$, 
i.e., $\partial_{x}T(m_{2y}>0)-\partial_{x}T(m_{2y}<0)=2 \partial_{x}T^{\rm ANE}(m_{2y}>0)$. 



\section{Conclusion}
\label{sec:Conclusion}

In conclusion, we developed a theory of spin transport in metallic ferromagnetic/nonmagnetic multilayer. 
We derived the theoretical formulas of the electric voltage via the inverse spin Hall effect, Eq. (\ref{eq:V_ISHE}), 
and the spin torque, Eq. (\ref{eq:STT}), excited through the anomalous Nernst effect 
by using a phenomenological equation of the spin-dependent current, Eq. (\ref{eq:current_def}). 
The estimated value of the temperature gradient which is necessary to obtain a large voltage or torque for practical application is 
at least one to two orders of magnitude larger than the experimentally available value using the current technique. 
The experimental efforts recently made, such as the material investigation \cite{sakuraba13,sakuraba16} 
and structure improvement \cite{uchida15}, however, will provide possibilities to observe these phenomena experimentally. 
The Seebeck and transverse spin Seebeck effects will also contribute to the experimental measurements. 
The contribution from the anomalous Nernst effect can be, however, evaluated separately from these effects 
by measuring the thickness and/or magnetization direction dependences of the generated voltage 
and the temperature gradient for the magnetization switching. 


\section*{Acknowledgement}

The author expresses gratitude to Mark D. Stiles, Wayne M. Saslow, Masamitsu Hayashi, and Yuya Sakuraba for having valuable discussions. 
In particular, Mark D. Stiles provided important comments to improve the manuscript. 
The author is also thankful to Shinji Yuasa, Kay Yakushiji, Hitoshi Kubota, Akio Fukushima, Takayuki Nozaki, Makoto Konoto, 
Sumito Tsunegi, Yoichi Shiota, Takehiko Yorozu, Satoshi Iba, Hiroki Maehara, and Ai Emura for their support and encouragement. 
This work was supported by JSPS KAKENHI Grant-in-Aid for Young Scientists (B) 16K17486. 



\appendix


\section{Diffusion equation of spin accumulation in the presence of nonuniform temperature gradient}
\label{sec:AppendixA}

Here, we derive the diffusion equation of the spin accumulation in the presence of nonuniform temperature gradient, 
i.e., $\bm{\nabla}^{2}T \neq 0$. 
Using Eq. (\ref{eq:electric_current_def}), 
the conservation law of the electric current, $\bm{\nabla}\cdot\mathbf{J}_{\rm c}=0$, gives 
\begin{equation}
  \bm{\nabla}^{2}
  \left(
    \bar{\mu}
    +
    \beta
    \delta
    \mu
  \right)
  =
  -e \mathscr{S}
  \bm{\nabla}^{2} T. 
\end{equation}
Note that the anomalous Nernst effect does not affect the conservation law of the electric current even in the presence of the nonuniform temperature gradient 
because of the vector formula, $\bm{\nabla}\cdot(\mathbf{m}\times\bm{\nabla}T)=0$. 
Then, the divergence of the spin current density, Eq. (\ref{eq:spin_current_def}), becomes 
\begin{equation}
  \bm{\nabla}
  \cdot
  \mathbf{J}_{\rm s}
  =
  -\frac{\hbar(1-\beta^{2})\sigma}{2e^{2}}
  \bm{\nabla}^{2}
  \delta
  \mu
  +
  \frac{\hbar (\beta-p_{\rm S}) \sigma}{2e}
  \mathscr{S}
  \bm{\nabla}^{2}T. 
\end{equation}
On the other hand, 
the divergence of the spin current density is \cite{valet93}, 
\begin{equation}
  \bm{\nabla}
  \cdot
  \mathbf{J}_{\rm s}
  =
  -\frac{\hbar (1-\beta^{2}) \sigma}{2e^{2} \ell^{2}}
  \delta 
  \mu.
\end{equation}
Then, we obtain the diffusion equation of the spin accumulation, 
\begin{equation}
  \bm{\nabla}^{2}
  \delta 
  \mu 
  =
  \frac{\delta\mu}{\ell^{2}}
  +
  e
  \frac{(\beta-p_{\rm S})}{(1-\beta^{2})}
  \mathscr{S}
  \bm{\nabla}^{2} T. 
  \label{eq:diffusion_equation_general}
\end{equation}
The last term, $e [(\beta-p_{\rm S})/(1-\beta^{2})] \mathscr{S} \bm{\nabla}^{2}T$, 
can be written as $-e (p_{\rm S}^{\prime}/2) S \bm{\nabla}^{2}T$, 
where $S=S_{\uparrow}+S_{\downarrow}$ and $p_{\rm S}^{\prime}=(S_{\uparrow}-S_{\downarrow})/(S_{\uparrow}+S_{\downarrow})$. 

The solution of the spin accumulation in Eq. (\ref{eq:diffusion_equation_general}) depends on the temperature profile. 
Note that the additional term to the spin diffusion equation is proportional to $\mathscr{S} \bm{\nabla}^{2}T$. 
Thus, the additional term can be neglected when we are only interested in the anomalous Nernst effect. 


\section{Explicit solutions of the coefficients in Eq. (\ref{eq:spin_accumulation_expand})}
\label{sec:AppendixB}

The conservation of the spin current in the nonmagnet, $\mathbf{I}_{s}^{\rm F_{1} \to N}+\mathbf{I}_{\rm s}^{\rm F_{2} \to N}=\bm{0}$, 
can be, in general, rewritten as 
\begin{equation}
\begin{split}
  &
  g_{\rm F_{1}}^{*}
  (\mathbf{m}_{1}\cdot\bm{\mu}_{\rm N})
  \mathbf{m}_{1}
  +
  g_{\rm r(F_{1})}
  \mathbf{m}_{1}
  \times
  \left(
    \bm{\mu}_{\rm N}
    \times
    \mathbf{m}_{1}
  \right)
  +
  g_{\rm i(F_{1})}
  \bm{\mu}_{\rm N}
  \times
  \mathbf{m}_{1}
\\
  &
  +
  g_{\rm F_{2}}^{*}
  (\mathbf{m}_{2}\cdot\bm{\mu}_{\rm N})
  \mathbf{m}_{2}
  +
  g_{\rm r(F_{2})}
  \mathbf{m}_{2}
  \times
  \left(
    \bm{\mu}_{\rm N}
    \times
    \mathbf{m}_{2}
  \right)
  +
  g_{\rm i(F_{2})}
  \bm{\mu}_{\rm N}
  \times
  \mathbf{m}_{2}
\\
  &=
  s_{1}
  \mathbf{m}_{1}
  -
  s_{2}
  \mathbf{m}_{2},
  \label{eq:equation_for_spin_accumulation_sub}
\end{split}
\end{equation}
where $s_{1}$ and $s_{2}$ are magnitudes of source terms. 
In the main text, only the F${}_{2}$ layer gives the source of spin current by the anomalous Nernst effect, 
and thus, $s_{1}=0$ and $s_{2}=-[\hbar g_{\rm F_{2}}^{*}/(2e g_{\rm F_{2}})]\tanh[d_{2}/(2\ell_{2})] (\beta-p_{\rm N}) \sigma_{\rm F_{2}}\mathscr{N} m_{2y} A \partial_{x}T$. 
For generality, let us assume that the F${}_{1}$ layer also provides source term. 
We note that, in the presence of source terms of transverse spin current, such as spin pumping \cite{tserkovnyak02a}, 
new source terms pointing in the direction perpendicular to the magnetization should be added to the right hand side. 
We expand $\bm{\mu}_{\rm N}$ as Eq. (\ref{eq:spin_accumulation_expand}), 
and introduce the notation $\tilde{z}=\mathbf{m}_{1}\cdot\mathbf{m}_{2}$. 
The exact solutions of $a$, $b$, and $c$ are given by 
\begin{equation}
\begin{split}
  a
  =&
  \frac{(a_{11} + a_{12} \tilde{z} + a_{13} \tilde{z}^{2} + a_{14} \tilde{z}^{3}) s_{1}}
    {a_{31} + a_{32} \tilde{z} + a_{33} \tilde{z}^{2} + a_{34} \tilde{z}^{3}}
\\
  &+
  \frac{(a_{21} + a_{22} \tilde{z} + a_{23} \tilde{z}^{2} + a_{24} \tilde{z}^{3}) s_{2}}
    {a_{31} + a_{32} \tilde{z} + a_{33} \tilde{z}^{2} + a_{34} \tilde{z}^{3}},
\end{split}
\end{equation}
where 
\begin{equation}
  a_{11}
  =
  -g_{\rm i(F_{1})} [ g_{\rm i(F_{1})}^{2} + (g_{\rm r(F_{1})} + g_{\rm F_{2}}^{*}) (g_{\rm r(F_{1})} + g_{\rm r(F_{2})}) ],
\end{equation}
\begin{equation}
  a_{12}
  =
  -g_{\rm i(F_{2})} [3 g_{\rm i(F_{1})}^{2} + (g_{\rm r(F_{1})} + g_{\rm F_{2}}^{*}) (g_{\rm r(F_{1})} + g_{\rm r(F_{2})}) ],
\end{equation}
\begin{equation}
  a_{13}
  =
  -g_{\rm i (F_{1})} [3 g_{\rm i(F_{2})}^{2} + (g_{\rm r(F_{2})} - g_{\rm F_{2}}^{*}) (g_{\rm r(F_{1})} + g_{\rm r(F_{2})}) ],
\end{equation}
\begin{equation}
  a_{14} 
  =
  -g_{\rm i(F_{2})} [g_{\rm i(F_{2})}^{2} + (g_{\rm r(F_{2})} - g_{\rm F_{2}}^{*}) (g_{\rm r(F_{1})} + g_{\rm r(F_{2})}) ],
\end{equation}
\begin{equation}
  a_{21}
  =
  g_{\rm i(F_{1})}^{2} g_{\rm i(F_{2})},
\end{equation}
\begin{equation}
  a_{22}
  =
  g_{\rm i(F_{1})} [g_{\rm i(F_{1})}^{2} + 2 g_{\rm i(F_{2})}^{2} + (g_{\rm r(F_{1})} + g_{\rm r(F_{2})})^{2} ],
\end{equation}
\begin{equation}
  a_{23}
  =
  g_{\rm i(F_{2})} [2 g_{\rm i(F_{1})}^{2} + g_{\rm i(F_{2})}^{2} + (g_{\rm r(F_{1})} + g_{\rm r(F_{2})})^{2} ],
\end{equation}
\begin{equation}
  a_{24}
  =
  g_{\rm i(F_{1})} g_{\rm i(F_{2})}^{2},
\end{equation}
\begin{equation}
\begin{split}
  a_{31}
  =&
  -g_{\rm i(F_{1})} 
  [ 
  g_{\rm F_{2}}^{*} g_{\rm i(F_{2})}^{2} + g_{\rm i(F_{2})}^{2} g_{\rm r(F_{1})} + g_{\rm i(F_{1})}^{2} (g_{\rm r(F_{2})} + g_{\rm F_{1}}^{*}) 
\\
  &+ g_{\rm F_{2}}^{*} (g_{\rm r(F_{2})} + g_{\rm F_{1}}^{*}) (g_{\rm r(F_{1})} + g_{\rm r(F_{2})}) 
\\
  &+ g_{\rm r(F_{1})} (g_{\rm r(F_{2})} + g_{\rm F_{1}}^{*}) (g_{\rm r(F_{1})} + g_{\rm r(F_{2})}) ],
\end{split}
\end{equation}
\begin{equation}
\begin{split}
  a_{32}
  =&
  -g_{\rm i(F_{2})} 
  [ 
  2 g_{\rm F_{2}}^{*} g_{\rm i(F_{1})}^{2} + g_{\rm i(F_{2})}^{2} g_{\rm F_{2}}^{*} + g_{\rm i(F_{2})}^{2} g_{\rm r(F_{1})} 
\\
  &- 2 g_{\rm i(F_{1})}^{2} g_{\rm r(F_{2})} + 3 g_{\rm i(F_{1})}^{2} ( g_{\rm F_{1}}^{*} + g_{\rm r(F_{2})}) 
\\
  &+ g_{\rm F_{2}}^{*} (g_{\rm r(F_{2})} + g_{\rm F_{1}}^{*}) (g_{\rm r(F_{1})} + g_{\rm r(F_{2})}) 
\\
  &+ g_{\rm r(F_{1})} (g_{\rm r(F_{2})} + g_{\rm F_{1}}^{*}) (g_{\rm r(F_{1})} + g_{\rm r(F_{2})}) ],
\end{split}
\end{equation}
\begin{equation}
\begin{split}
  a_{33}
  =&
  g_{\rm i(F_{1})} 
  \{ 
  g_{\rm i(F_{2})}^{2} g_{\rm r(F_{1})} + g_{\rm i(F_{1})}^{2} g_{\rm r(F_{2})} 
\\
  &+ g_{\rm r(F_{1})} g_{\rm r(F_{2})} (g_{\rm r(F_{1})} + g_{\rm r(F_{2})}) 
\\
  & - g_{\rm F_{2}}^{*} [ g_{\rm i(F_{1})}^{2} + 2 g_{\rm i(F_{2})}^{2} + (g_{\rm r(F_{1})} - g_{\rm F_{1}}^{*}) (g_{\rm r(F_{1})} + g_{\rm r(F_{2})}) ] 
\\
  &- g_{\rm F_{1}}^{*} [3 g_{\rm i(F_{2})}^{2} + g_{\rm r(F_{2})} (g_{\rm r(F_{1})} + g_{\rm r(F_{2})}) ]\},
\end{split}
\end{equation}
\begin{equation}
\begin{split}
  a_{34}
  =&
  g_{\rm i(F_{2})} 
  \{ 
  g_{\rm i(F_{2})}^{2} g_{\rm r(F_{1})} + g_{\rm i(F_{1})}^{2} g_{\rm r(F_{2})} 
\\
  &+ g_{\rm r(F_{1})} g_{\rm r(F_{2})} (g_{\rm r(F_{1})} + g_{\rm r(F_{2})}) 
\\
  &- g_{\rm F_{2}}^{*} [g_{\rm i(F_{1})}^{2} + (g_{\rm r(F_{1})} - g_{\rm F_{1}}^{*}) (g_{\rm r(F_{1})} + g_{\rm r(F_{2})}) ] 
\\
  &- g_{\rm F_{1}}^{*} [g_{\rm i(F_{2})}^{2} + g_{\rm r(F_{2})} (g_{\rm r(F_{1})} + g_{\rm r(F_{2})}) ] \},
\end{split}
\end{equation}
and 
\begin{equation}
  b
  =
  \frac{(b_{11} + b_{12} \tilde{z}) s_{1} + (b_{21} + b_{22} \tilde{z}) s_{2}}{b_{31} + b_{32} \tilde{z} + b_{33} \tilde{z}^{2}},
\end{equation}
where
\begin{equation}
  b_{11}
  =
  g_{\rm i(F_{2})} (g_{\rm r(F_{1})} + g_{\rm F_{2}}^{*}), 
\end{equation}
\begin{equation}
  b_{12}
  =
  -g_{\rm i(F_{1})} (g_{\rm r(F_{2})} - g_{\rm F_{2}}^{*}),
\end{equation}
\begin{equation}
  b_{21}
  =
  g_{\rm i(F_{1})} (g_{\rm r(F_{2})} + g_{\rm F_{1}}^{*}), 
\end{equation}
\begin{equation}
  b_{22}
  =
  -g_{\rm i(F_{2})} (g_{\rm r(F_{1})} - g_{\rm F_{1}}^{*}),
\end{equation}
\begin{equation}
\begin{split} 
  b_{31}
  =&
  -g_{\rm i(F_{2})}^{2} g_{\rm r(F_{1})} - g_{\rm i(F_{1})}^{2} g_{\rm r(F_{2})} 
\\
  &- g_{\rm r(F_{1})} g_{\rm r(F_{2})} (g_{\rm r(F_{1})} + g_{\rm r(F_{2})}) 
\\
  &- g_{\rm F_{1}}^{*} [g_{\rm i(F_{1})}^{2} + (g_{\rm r(F_{1})} + g_{\rm F_{2}}^{*}) (g_{\rm r(F_{1})} + g_{\rm r(F_{2})})] 
\\
  &- g_{\rm F_{2}}^{*} [g_{\rm i(F_{2})}^{2} + g_{\rm r(F_{2})} (g_{\rm r(F_{1})} + g_{\rm r(F_{2})}) ],
\end{split}
\end{equation}
\begin{equation}
  b_{32}
  =
  -2 g_{\rm i(F_{1})} g_{\rm i(F_{2})} (g_{\rm F_{1}}^{*} + g_{\rm F_{2}}^{*}),
\end{equation} 
\begin{equation}
\begin{split}
  b_{33}
  =&
  g_{\rm i(F_{2})}^{2} g_{\rm r(F_{1})} + g_{\rm i(F_{1})}^{2} g_{\rm r(F_{2})} 
\\
  &+ g_{\rm r(F_{1})} g_{\rm r(F_{2})} (g_{\rm r(F_{1})} + g_{\rm r(F_{2})}) 
\\
  &- g_{\rm F_{1}}^{*} [g_{\rm i(F_{2})}^{2} + g_{\rm r(F_{2})} (g_{\rm r(F_{1})} + g_{\rm r(F_{2})})] 
\\
  & - g_{\rm F_{2}}^{*} [g_{\rm i(F_{1})}^{2} + (g_{\rm r(F_{1})} - g_{\rm F_{1}}^{*}) (g_{\rm r(F_{1})} + g_{\rm r(F_{2})}) ], 
\end{split}
\end{equation}
and 
\begin{equation}
  c
  =
  \frac{(c_{11} + c_{12} \tilde{z}) s_{1} + (c_{21} + c_{22} \tilde{z}) s_{2}}{c_{31} + c_{32} \tilde{z} + c_{33} \tilde{z}^{2}}
\end{equation}
where 
\begin{equation}
  c_{11}
  =
  -g_{\rm i(F_{1})} g_{\rm i(F_{2})} ,
\end{equation}
\begin{equation}
  c_{12}
  =
  -g_{\rm i(F_{2})}^{2} - (g_{\rm r(F_{2})} - g_{\rm F_{2}}^{*}) (g_{\rm r(F_{1})} + g_{\rm r(F_{2})}),
\end{equation}
\begin{equation}
  c_{21}
  =
  g_{\rm i(F_{2})}^{2} + (g_{\rm r(F_{2})} + g_{\rm F_{1}}^{*}) (g_{\rm r(F_{1})} + g_{\rm r(F_{2})}),
\end{equation}
\begin{equation}
  c_{22}
  =
  g_{\rm i(F_{1})} g_{\rm i(F_{2})},
\end{equation}
\begin{equation}
\begin{split}
  c_{31}
  =&
  -g_{\rm i(F_{2})}^{2} g_{\rm r(F_{1})} - g_{\rm i(F_{1})}^{2} g_{\rm r(F_{2})} 
\\
  &- g_{\rm r(F_{1})} g_{\rm r(F_{2})} (g_{\rm r(F_{1})} + g_{\rm r(F_{2})}) 
\\
  &- g_{\rm F_{1}}^{*} [g_{\rm i(F_{1})}^{2} + (g_{\rm r(F_{1})} + g_{\rm F_{2}}^{*}) (g_{\rm r(F_{1})} + g_{\rm r(F_{2})})] 
\\
  &- g_{\rm F_{2}}^{*} [g_{\rm i(F_{2})}^{2} + g_{\rm r(F_{2})} (g_{\rm r(F_{1})} + g_{\rm r(F_{2})})],
\end{split}
\end{equation}
\begin{equation}
  c_{32}
  =
  -2 g_{\rm i(F_{1})} g_{\rm i(F_{2})} (g_{\rm F_{1}}^{*} + g_{\rm F_{2}}^{*}),
\end{equation}
\begin{equation}
\begin{split}
  c_{33}
  =&
  g_{\rm i(F_{2})}^{2} g_{\rm r(F_{1})} + g_{\rm i(F_{1})}^{2} g_{\rm r(F_{2})} 
\\
  &+ g_{\rm r(F_{1})} g_{\rm r(F_{2})} (g_{\rm r(F_{1})} + g_{\rm r(F_{2})}) 
\\
  &- g_{\rm F_{1}}^{*} [g_{\rm i(F_{2})}^{2} +  g_{\rm r(F_{2})} (g_{\rm r(F_{1})} + g_{\rm r(F_{2})})] 
\\
  &- g_{\rm F_{2}}^{*} [g_{\rm i(F_{1})}^{2} + (g_{\rm r(F_{1})} - g_{\rm F_{1}}^{*}) (g_{\rm r(F_{1})} + g_{\rm r(F_{2})})].
\end{split}
\end{equation}
The spin torque acting on the magnetization of the F${}_{1}$ layer is 
\begin{equation}
\begin{split}
  \mathbf{T}
  &=
  \frac{\gamma_{0}}{M_{1}V_{1}}
  \mathbf{m}_{1}
  \times
  \left(
    \mathbf{I}_{\rm s}^{\rm F_{1} \to N}
    \times
    \mathbf{m}_{1}
  \right)
\\
  &=
  -\frac{\gamma_{0}}{4\pi M_{1}V_{1}}
  \left[
    \left(
      g_{\rm r(F_{1})} c
      +
      g_{\rm i(F_{1})} b
    \right)
    \mathbf{m}_{1}
    \times
    \left(
      \mathbf{m}_{2}
      \times
      \mathbf{m}_{1}
    \right)
  \right.
\\
  &
  \left.
    +
    \left(
      g_{\rm r(F_{1})} b
      -
      g_{\rm i(F_{1})} c
    \right)
    \mathbf{m}_{1}
    \times
    \mathbf{m}_{2}
  \right].
\end{split}
\end{equation}
As shown, the coefficient $a$ does not appear in the spin torque formula. 
This is because only the transverse (normal to the magnetization) component of the spin accumulation provides spin torque. 
When the imaginary part of the mixing conductance is negligibly small ($g_{\rm i} \to 0$) \cite{brataas06}, 
the coefficient $b$ becomes zero.




\begin{thebibliography}{99}
\bibitem{baibich88} M. N. Baibich, J. M. Broto, A. Fert, F. N. van Dau, F. Petroff, P. Etienne, G. Creuzet, A. Friederich, and J. Chazelas:
                    Phys. Rev. Lett. $\bm{61}$, 2472 (1988).
\bibitem{binasch89} G. Binasch, P. Gr\"unberg, F. Saurenbach, and W. Zinn: Phys. Rev. B $\bm{39}$, 4828 (1989). 
\bibitem{pratt91} W. P. Pratt, S.-F. Lee, J. M. Slaughter, R. Loloee, P. A. Schroeder, and J. Bass: Phys. Rev. Lett. $\bm{66}$, 3060 (1991). 
\bibitem{slonczewski96} J. C. Slonczewski: J. Magn. Magn. Mater. $\bm{159}$, L1 (1996). 
\bibitem{berger96} L. Berger: Phys. Rev. B $\bm{54}$, 9353 (1996). 
\bibitem{johnson88} M. Johnson and R. H. Silsbee: Phys. Rev. Lett. $\bm{60}$, 377 (1988).
\bibitem{johnson93} M. Johnson: Phys. Rev. Lett. $\bm{70}$, 2142 (1993). 
\bibitem{jedema03} F. J. Jedema, A. T. Filip, and B. J. van Wees: Phys. Rev. B $\bm{67}$, 085319 (2003). 
\bibitem{kimura06} T. Kimura, Y. Otani, and J. Hamrle: Phys. Rev. Lett. $\bm{96}$, 037201 (2006). 
\bibitem{silsbee79} R. H. Silsbee, A. Janossy, and P. Monod: Phys. Rev. B $\bm{19}$, 4382 (1979).
\bibitem{mizukami02a} S. Mizukami, Y. Ando, and T. Miyazaki: J. Magn. Magn. Mater. $\bm{239}$, 42 (2002). 
\bibitem{mizukami02b} S. Mizukami, Y. Ando, and T. Miyazaki: Phys. Rev. B $\bm{66}$, 104413 (2002). 
\bibitem{tserkovnyak02a} Y. Tserkovnyak, A. Brataas, and G. E. W. Bauer: Phys. Rev. Lett. $\bm{88}$, 117601 (2002). 
\bibitem{tserkovnyak02b} Y. Tserkovnyak, A. Brataas, and G. E. W. Bauer: Phys. Rev. B $\bm{66}$, 224403 (2002). 
\bibitem{dyakonov71} M. I. Dyakonov and V. I. Perel: Phys. Lett. A $\bm{35}$, 459 (1971). 
\bibitem{hirsch99} J. E. Hirsch: Phys. Rev. Lett. $\bm{83}$, 1834 (1999). 
\bibitem{kato04} Y. K. Kato, R. C. Myers, A. C. Gossard, and D. D. Awschalom: Science $\bm{306}$, 1910 (2004). 
\bibitem{ando08} K. Ando, S. Takahashi, K. Harii, K. Sasage, J. Ieda, S. Maekawa, and E. Saitoh: Phys. Rev. Lett. $\bm{101}$, 036601 (2008). 
\bibitem{uchida08} K. Uchida, S. Takahashi, K. Harii, J. Ieda, W. Koshibae, K. Ando, S. Maekawa, and E. Saitoh: Nature $\bm{455}$, 778 (2008). 
\bibitem{uchida10} K. Uchida, J. Xiao, H. Adachi, J. Ohe, S. Takahashi, J. Ieda, T. Ota, Y. Kajiwara, H. Umezawa, H. Kawai, G. E. W. Bauer, S. Maekawa, and E. Saitoh: Nat. Mater. $\bm{9}$, 894 (2010). 
\bibitem{slachter10} A. Slachter, F. L. Bakker, J.-P. Adam, and B. J. van Wees: Nat. Phys. $\bm{6}$, 879 (2010). 
\bibitem{bauer12} G. E. W. Bauer, E. Saitoh, and B. J. van Wees: Nat. Mater. $\bm{11}$, 391 (2012). 
\bibitem{takahashi16} R. Takahashi, M. Matsuo, M. Ono, K. Harii, H. Chudo, S. Okayasu, J. Ieda, S. Takahashi, S. Maekawa, and E. Saitoh: Nat. Phys. $\bm{12}$, 52 (2016). 
\bibitem{huang11} S. Y. Huang, W. G. Wang, S. F. Lee, J. Kwo, and C. L. Chien: Phys. Rev. Lett. $\bm{107}$, 216604 (2011). 
\bibitem{sakuraba13} Y. Sakuraba, K. Hasegawa, M. Mizuguchi, T. Kubota, S. Mizukami, T. Miyazaki, and K. Takanashi: Appl. Phys. Express $\bm{6}$, 033003 (2013). 
\bibitem{sakuraba16} Y. Sakuraba: Scr. Mater. $\bm{111}$, 29 (2016). 
\bibitem{uchida15} K. Uchida, T. Kikkawa, T. Seki, T. Oyake, J. Shimoi, Z. Qiu, K. Takanashi, and E. Saitoh: Phys. Rev. B $\bm{92}$, 094414 (2015). 
\bibitem{hatami07} M. Hatami, G. E. W. Bauer, Q. Zhang, and P. J. Kelly: Phys. Rev. Lett. $\bm{99}$, 066603 (2007). 
\bibitem{hatami09} M. Hatami, G. E. W. Bauer, Q. Zhang, and P. J. Kelly: Phys. Rev. B $\bm{79}$, 174426 (2009). 
\bibitem{xiao10} J. Xiao, G. E. W. Bauer, K. Uchida, E. Saitoh, and S. Maekawa: Phys. Rev. B $\bm{81}$, 214418 (2010). 
\bibitem{adachi10} H. Adachi, K. Uchida, E. Saitoh, J. Ohe, S. Takahashi, and S. Maekawa: Appl. Phys. Lett. $\bm{97}$, 252506 (2010). 
\bibitem{adachi11} H. Adachi, J. Ohe, S. Takahashi, and S. Maekawa: Phys. Rev. B $\bm{83}$, 094410 (2011). 
\bibitem{taniguchi15} T. Taniguchi, J. Grollier, and M. D. Stiles: Phys. Rev. Applied $\bm{3}$, 044001 (2015). 
\bibitem{valet93} T. Valet and A. Fert: Phys. Rev. B $\bm{48}$, 7099 (1993). 
\bibitem{brataas01} A. Brataas, Y. V. Nazarov, and G. E. W. Bauer: Eur. Phys. J. B $\bm{22}$, 99 (2001). 
\bibitem{taniguchi08} T. Taniguchi, S. Yakata, H. Imamura, and Y. Ando: Appl. Phys. Express $\bm{1}$, 031302 (2008). 
\bibitem{ghosh12} A. Ghosh, S. Auffret, U. Ebels, and W. E. Bailey: Phys. Rev. Lett. $\bm{109}$, 127202 (2012). 
\bibitem{scharf12} B. Scharf, A. Motos-Abiague, I. Zutic, and J. Fabian: Phys. Rev. B $\bm{85}$, 085208 (2012). 
\bibitem{schmid13} M. Schmid, S. Srichandan, D. Meier, T. Kuschel, J.-M. Schamalhorst, M. Vogel, G. Reiss, C. Strunk, and C. H. Back: Phys. Rev. Lett. $\bm{111}$, 187201 (2013). 
\bibitem{kikkawa13} T. Kikkwa, K. Uchida, Y. Shiomi, Z. Qiu, D. Hou, D. Tian, H. Nakayama, X.-F. Jin, and E. Saitoh: Phys. Rev. Lett. $\bm{110}$, 067207 (2013). 
\bibitem{bass07} J. Bass and J. W. P. Pratt Jr: J. Phys.: Condens. Matter $\bm{19}$, 183201 (2007). 
\bibitem{brataas06} A. Brataas, G. E. W. Bauer, and P. J. Kelly: Phys. Rep. $\bm{427}$, 157 (2006). 
\bibitem{uchida09} K. Uchida, S. Takahashi, J. Ieda, K. Harii, K. Ikeda, W. Koshibae, S. Meakawa, and E. Saitoh: J. Appl. Phys. $\bm{105}$, 07C908 (2009). 
\bibitem{isasa15} M. Isasa, E. Villamor, L. E. Hueso, M. Gradhand, and F. Casanova: Phys. Rev. B $\bm{91}$, 024402 (2015). 
\bibitem{liu12} L. Liu, C.-F. Pai, Y. Li, H. W. Tseng, D. C. Ralph, and R. A. Buhrman: Science $\bm{336}$, 555 (2012). 
\bibitem{jia11} C. Jia and J. Berakdar: Phys. Rev. B $\bm{83}$, 180401 (2011). 
\bibitem{tsunegi14} S. Tsunegi, H. Kubota, S. Tamaru, K. Yakushiji, M. Konoto, A. Fukushima, T. Taniguchi, H. Arai, H. Imamura, and S. Yuasa: Appl. Phys. Express $\bm{7}$, 033004 (2014). 
\bibitem{uchida10s} K. Uchida, T. Ota, K. Harii, S. Takahashi, S. Maekawa, Y. Fujiwara, and E. Saitoh: Solid State Commun. $\bm{150}$, 524 (2010). 
\end{thebibliography}


\end{document}